\documentclass[11pt]{article}
\usepackage{moriond, amsmath}
\usepackage{graphicx, subfig}

\def\Journal#1#2#3#4{{#1} {\bf #2}, #3 (#4)}
\def\PRL{\em Phys. Rev. Lett.}
\def\PRD{{\em Phys. Rev.} D}

\begin{document}
\vspace*{4cm}
\title{Studying Neutrino Directionality with Double Chooz}

\author{ E. Caden }

\address{Department of Physics, Drexel University,\\
Philadelphia, PA, 19104, USA}

\maketitle\abstracts{The first results from Double Chooz with 100 days of data measured $\sin^2(2\theta_{13})
= 0.086 \pm 0.051$. Backgrounds contribute significantly to the systematic uncertainty budget. Using the
incoming neutrino directionality we will attempt to further separate the inverse beta decay signal from
backgrounds. The CHOOZ experiment completed a similar analysis and found that the neutrino source can be
located to within a cone of half-aperture of 18 degrees at the 68\% C.L. We study the possible improvement
of this result by Double Chooz.}

\section{Double Chooz}
\label{sec:DC}
Double Chooz is a neutrino oscillation experiment in the French Ardennes. It is a reactor experiment, looking
for a deficit of electron neutrinos.  The probability that an electron anti-neutrino will be detected as an
electron anti-neutrino is 
\begin{equation}
\label{eq:Surv}
P_{surv}\approx1-\sin^2 2\theta_{13}\sin^2(1.267\Delta m^2 L/E),
\end{equation}
where $\Delta m^2$ is the atmospheric mass splitting, $L$ is the distance from the reactor cores to the
detector in meters, and $E$ is the energy of the anti-neutrino in MeV. Electron anti-neutrinos are produced by
two 4.25~GW$_{TH}$ reactor cores of the Chooz Nuclear Power Plant. Neutrinos are detected through the inverse
beta decay reaction (IBD):
\begin{equation}
\label{eq:IBD}
 \bar{\nu}_e+p\rightarrow n+e^+
\end{equation}

\begin{figure}[h!]
\subfloat[The Double Chooz detector system.] {\label{fig:DCdet}
\includegraphics[width=0.5\textwidth]{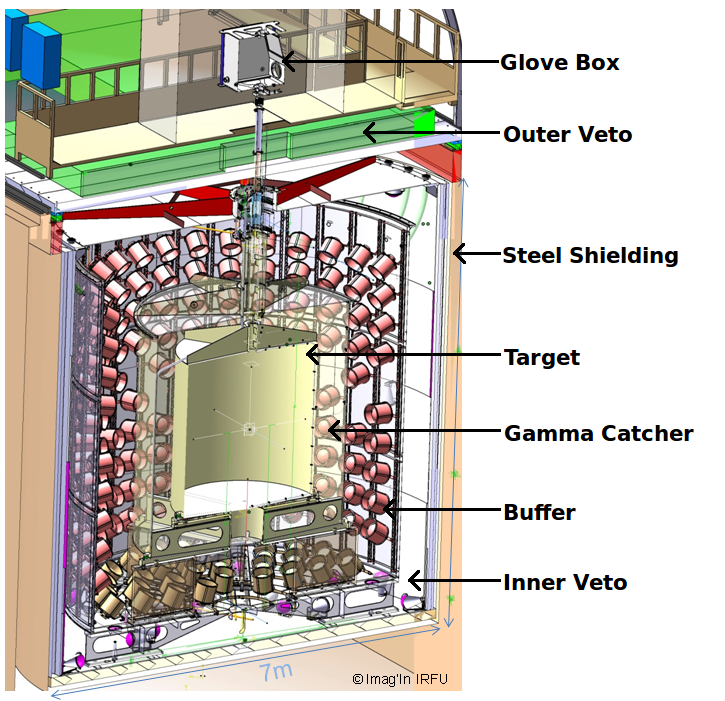}}
\subfloat[Top: Measured and expected prompt energy spectra, including backgrounds. Bottom:
Difference between data and no oscillation spectrum.] {\label{fig:DCResult}
\includegraphics[width=0.5\textwidth]{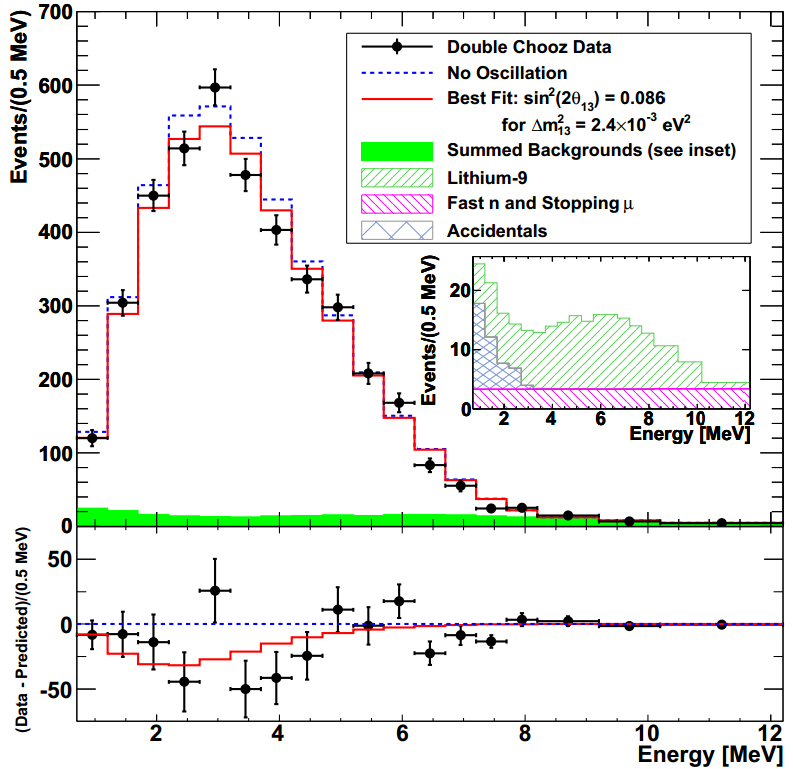}}
\end{figure}

The detector (fig.~\ref{fig:DCdet}), $1.05$~km from the reactors, consists of a series of sub-detectors: the
neutrino target, a
gamma catcher region, a non-scintillating buffer, an inner veto region, and an outer veto cover. The
innermost vessel is target, an acrylic vessel filled with $10.3$~m$^3$ of PXE and dodecane with fluors PPO and
Bis-MSB. The target also contains Gadolinium at $0.1\%$ concentration, to enhance the neutron capture
probability. The gamma catcher is also filled with scintillating mineral oil, and is used to detect gammas
escaping from the target. There are 390 photomultiplier tubes (PMTs) mounted in the buffer, which is
filled with non scintillating oil. This region shields the target from radioactivity of the photomultiplier
tubes. The inner veto, with 78 PMTs and filled with scintillator, vetoes cosmic-ray muons and spallation
neutrons. The entire volume is then surrounded by low activity demagnetized steel. To also help in tagging
muons, Double Chooz has an outer veto, which is sixty-four panels of scintillator strips and wavelength
shifting fibers that tag the position and angle of incoming muons. The entire detector system is buried under
300~m.w.e of rock overburden, which acts as a natural shield against cosmic ray muons.

The signals are amplified by an analog Front End system, and events are triggered based on the summed charge
of PMT groups exceeding an energy threshold. When triggered, waveforms from each PMT are digitized by
Flash-ADC electronics with a custom, dead-time free acquisition system. After 100 days of physics data, 4121
neutrino candidates were found, when $4344\pm165$ were expected with no oscillation. This deficit leads to
$\sin^22\theta_{13} = 0.086 \pm 0.051$ (stat + sys)\cite{DC_FirstResult_2012}. The result is shown in
figure~\ref{fig:DCResult}.

\section{Neutrino Directionality}
\label{sec:Directionality}
Measuring neutrino directionality depends on accurately knowing the position and angular distribution of the
IBD reaction products: the positron and the neutron. Results presented here are relevant for anti-neutrinos
produced by reactors, $E_{\nu} < 10$~Mev.

From Vogel and Beacom~\cite{Vogel_angular_1999}, the positron angular distribution is given by
\begin{equation}
 \frac{d\sigma}{d\cos\theta}\simeq 1+\beta a(E_{\nu})\cos\theta_e
\end{equation}
where $\theta_e$ is the angle between anti-neutrino and positron directions, $\beta$ is the positron
velocity, and $a$ is the asymmetry coefficient. $a(E_{\nu})$ is given by the competition of non-spin-flip
(Fermi) and spin-flip (Gamow-Teller) contributions. In the limit where the mass of the nucleon is taken to be
infinite $(1/m_n\approx0)$, $a(E_{\nu})= a^{(0)}$,
\begin{equation}
 a^{(0)}=\frac{f^2-g^2}{f^2+3g^2}\simeq-0.102
\end{equation}
where $f$ and $g$ are the vector and axial-vector coupling constants, respectively taken to be 1 and 1.26.
Taking the common assumption of relativistic neutrinos where $\beta\approx 1$, the average cosine weighted
by the differential cross section is
\begin{equation}
\left<\cos\theta_e\right>=\frac{1}{3} a(E_{\nu})=-0.034.
\end{equation}
The average positron angle is slightly backwards. The original CHOOZ collaboration calculated the mean
positron displacement to be $-0.05 $~cm with respect to the initial anti-neutrino interaction
point~\cite{CHOOZ_directionality_1999}. The small displacement and negligible preferred scattering
angle mean that the positron movement can be neglected. Accounting for the small kinetic energy of the
neutron, the
positron's energy depends on its scattering angle and is
\begin{equation}
 E_e=(E_{\nu}-\Delta)\left[1-\frac{E_{\nu}}{m_n}(1-\cos\theta_e)\right]-\frac{\Delta^2-m_e^2}{2m_n}
\end{equation}
The neutron's kinetic energy then follows as
\begin{equation}
 E_n=\frac{E_{\nu}\left(E_{\nu}-\Delta\right)}{m_n}(1-\cos\theta_e)+\frac{\Delta^2-m_e^2}{2m_n}
\end{equation}
%{\bf [plot of positron and neutron Energy as function of neutrino energy]}
Conservation of momentum in the laboratory ($\vec{p}_p=0$) leads to 
\begin{equation}
 \vec{p}_{\nu}=\vec{p}_{e}+\vec{p}_{n}.
\end{equation}
Using
\begin{equation}
 |p_e| \leq\sqrt{(E_{\nu}-\Delta)^2-m_e^2} < E_{\nu},
\end{equation}
one can calculate that the neutron must be always be emitted in the forward hemisphere, relative to
the incoming anti-neutrino direction. The maximum angle between the anti-neutrino and initial neutron
directions was calculated in \cite{Vogel_angular_1999} to be
\begin{equation}
 \cos(\theta_n)_{max}=\frac{\sqrt{2E_{\nu}\Delta-(\Delta^2-m_e^2)}}{E_{\nu}}.
\end{equation}
%{\bf[plot of neutron momentum, line of max and sample of simulated]}
At the IBD threshold of 1.8MeV the neutron direction is purely forward and for reactor anti-neutrino
energies ($<10$~MeV) it is still mainly in the forward direction. Neutron diffusion preserves the incoming
direction of the neutron. With each scatter, the average cosine with respect to the incoming direction is:
\begin{equation}
 <\cos\theta_n> = 2/3A
\end{equation}
where $A$ is the atomic number of the scattering nucleus. Therefore the direction is most preserved when
scattering off Hydrogen, which has a large cross-section for neutron scattering at the relevant energy range.

Reconstructing the incoming neutrino's direction requires constructing a vector, $\vec{X}^i_{en}$, pointing
from the positron's reconstructed position to the neutron's reconstructed position, where
$\vec{X}^i_{en}=X^i_n-X^i_e$. Averaging this vector over all events gives:
\begin{equation}
 \vec{p}=\sum_{i=i}^N \vec{X}^i_{en}
\end{equation}
The cosine of the angle between $\vec{X}^i_{en}$ and a similar vector from the reactor to the detector
$\vec{X}_{RD}$ should point more towards +1 than -1. This is a statistical measurement that improves
with the number of neutrino candidates. The probability distribution of the components of $\vec{p}$
are Gaussian, with a width $P$. Because of the symmetry of the Double Chooz detector's cylindrical
geometry, $P$ is expected to be the same for all directions. Measuring $\vec{p}=(p_x, p_y, p_z)$, we can
calculate the azimuthal and zenith angles of the neutrino source:
\begin{eqnarray*}
 \tan\phi&=&\frac{\bar{p}_y}{\bar{p}_x}\\
\tan\theta&=&\frac{\bar{p}_z}{\sqrt{\bar{p}^2_x+\bar{p}^2_y}}
\end{eqnarray*}
The possibility of reconstructing the direction of the neutrino source depends on the errors on the angles,
$\Delta\phi$ and $\Delta\theta$. Using the gaussian nature of the components, $\Delta p_i=P/\sqrt{N}$, we
calculate the errors to be:
\begin{eqnarray*}
 \Delta\tan\phi&=&\sqrt{\left(\frac{P\cdot\bar{p}_y}{\sqrt{N}\bar{p}^2_x}\right)^2+\left(\frac{P}{\sqrt{N}\bar
{ p } _x } \right)^2 } \\
 \Delta\tan\theta&=&\sqrt{\left(\frac{P}{\sqrt{N}\sqrt{\bar{p}^2_x+\bar{p}^2_y}}\right)^2
\left[1+\left(\frac{\bar{p}_y \bar{p}_z}{\bar{p}_x^2+\bar{p}_y^2}\right)^2+\left(\frac{\bar{p}_x
\bar{p}_z}{\bar{p}_x^2+\bar{p}_y^2}\right)^2 \right]}
\end{eqnarray*}

\section{Previous work on Directionality and Double Chooz Improvement}
\label{sec:DCImprovement}
The concept of reactor neutrino directionality was shown by the G\"osgen, Bugey, and Palo Verde collaborations
for segmented scintillator detectors. These measurements were made by observing the neutron in a segment
farther away from the reactor than the segment where the positron was observed.

The first measurement of this kind using an unsegmented scintillator detector was by the CHOOZ collaboration,
using the technique described above. With only $\sim2500$ events, the neutrino source was located within a
cone of $18^{\circ}$~half-angle opening. As this is primarily a statistics driven measurement, Double Chooz
is poised to greatly improve that result.

\end{document}